\documentstyle[12pt]{article}

\newcommand{\be}{\begin{equation}}
\newcommand{\ee}{\end{equation}}
\newcommand{\ba}{\begin{array}}
\newcommand{\ea}{\end{array}}
\newcommand{\bc}{\begin{center}}
\newcommand{\ec}{\end{center}}
\newcommand{\disregard}[1]{{}}
\newcommand{\ti}{\tilde}

\newcommand{\nc}{n_c}
\newcommand{\nt}{\ti{n}}
\newcommand{\mt}{\ti{m}}

\newcommand{\ds}{\displaystyle}

\newcommand{\demi}{{\ds 1\over\ds 2}}

\begin{document}

\title{Anomalous Effects in a Trapped Bose-Einstein Condensate}
\author{M. Benarous \\
{\it Laboratory for Theoretical Physics and Material Physics} \\
{\it Faculty of Sciences} \\
{\it Hassiba Benbouali University of Chlef (Algeria)} }

\date{\today}
\maketitle

\begin{abstract}

We study the anomalous density in an ultra-cold trapped bose gas in a variational framework, for both zero and finite temperature. 
We show that it is finite in 1D, while it is logarithmically and linearly divergent in 2D and 3D. The renormalization that we adopt 
is more reliable and compatible with the variational scheme. The main outcome is that the anomalous and non condensate densities are 
of the same order of magnitude and should therefore be treated on an equal footing.

\end{abstract}

PACS: 05.30.Jp, 11.15.Tk, 32.80.Pj

\newpage

\setcounter{equation}{0}

\section{Introduction}

In a previous paper \cite{BM05}, using the time dependent variational principle of Balian and V\'en\'eroni\cite{BV}, we have derived variationally a set of dynamical coupled equations (which we called TDHFB for Time-Dependent Hartree-Fock-Bogoliubov) for dilute trapped bose gases 
below the transition. These equations were shown to generalize in a consistent way the Gross-Pitaevskii equations\cite{GP} in that they introduce a non trivial coupling between the order parameter, the thermal cloud and the anomalous density.

In order to apprehend better this coupling and to understand the role of the anomalous density, we extended the formalism to the non-local case\cite{BB10, BB11}, thus obtaining an intrinsic dynamics of the thermal cloud and the anomalous average which has never been written down before (an exception is perhaps the paper of Chernyak {\it et al.} \cite{CHER03} which discusses a somewhat similar set of equations using the generalized coherent state representation). We focused in these last two papers on the anomalous density owing to its importance to account for many-body effects. Nonetheless, the subsequent analysis was mainly performed in the low momentum regime, therefore leading to an effective theory which takes into account only the low lying modes. In this respect, the divergences of the anomalous density or the many-body T-matrix which are known to occur at high momenta were hidden in our treatment.

Despite the fact that the resulting equations were greatly simplified in the static case, leading to useful conclusions about the anomalous density, it was not obvious that the inclusion of higher modes will maintain these results, especially at finite temperature. In particular, it was observed \cite{MOR} that the coupled dynamics of the atoms in and out of the condensate induces an effect which is highly needed in order to go beyond the HFB-Popov approximations. Indeed, the discrepancies between the HFB-Popov computed properties and the experimental resuts that show up at temperatures as high as 60\% of the transition temperature seem to be related to the absence of a feedback effect of the dynamics of both the thermal cloud and the anomalous average on the condensate density\cite{HUTCH}.

In this paper, we would like to go beyond the approximations discussed in \cite{BB10, BB11} by explicitely taking into account the high lying modes. This will obviously lead to self-consistent equations exhibiting UV divergences that have to be regularized.

In this context, we note that in reference \cite{CHER03}, the diverging terms misteriously disappear from the final equations. The subsequent analysis performed there is thus of a very limited validity. In \cite{Pr11} however, the $\Lambda$ potential technique\cite{Pr04} is developed in order to obtain a consistent HFB theory. This formalism is however not fully variational and endows many intricacies such as the parameter $\Lambda$ itself. In our case, we will use instead the Wigner representation\cite{TT97, RS80, SV00} which is best suited for nonuniform systems and which will allow us to handle the divergences in a variational framework. We will see that we arrive at the same renormalization scheme as \cite{Pr11} in a more reliable way. Moreover, we obtain important quantitative estimates for the anomalous average and the non condensate density leading to the fact that they are essentially of the same order of magnitude. We therefore arrive at the conclusion that what is called in the literature the HFB-Popov approximation\cite{MOR, HUTCH, GRIF96, PROUK96} is a quite hazardous and highly inconsistent approximation\cite{YUK}.

The paper is organized as follows. In section 2, we recall the main steps in the variational derivation of the time-dependent Hartree-Fock-Bogoliubov equations for a system of trapped bosons with general two-body interaction and study some of their properties. Then we specialize to the contact potential case which allows us to show that the HFB-Bogoliubov-De Gennes (HFB-BdG) equations are just the quasi-homogeneous limit of our general equations.
In section 3, we discuss the zero temperature static solutions. We focus in particular on the anomalous density owing to its importance to account for many-body effects. We derive analytic expressions for both the anomalous and the noncondensate densities for one, two and three dimensional systems.
The anomalous density is shown to be finite for $D=1$ while it is logarithmically and linearly divergent for $D=2$ and 3. Our renormalization scheme is 
not only more reliable and consistent with the variational approach, but it leads to the same condition of renormalization found with the pseudo-potential methods. Most importantly, the anomalous and the non condensate densities, although small at zero temperature, are found to be of the same order of magnitude, leading therefore to the conclusion that it would be very hazardous to neglect one of them while keeping the other such as what is performed in the so (wrongly) called  Popov approximation.

In section 4, we evaluate the leading terms in the low temperature approximation and observe that our conclusions are not altered by the finite temperature corrections. Finaly, we present some concluding remarks and perspectives.

\setcounter{equation}{0}

\section{The TDHFB Equations}

Let us first recall how the generalized TDHFB equations may be derived in a variational framework. The Balian-V\'en\'eroni variational principle requires the choice of a trial space for the density operator. We will consider a gaussian time-dependent density-like operator. This Ansatz, which belongs to the class of the generalized coherent states, allows us to perform the calculation (since there exists a Wick's theorem) while retaining some fundamental aspects such as the pairing between atoms. Let us first define the quantities
\be
\ba{rl}
\Phi ({\bf r}) & = \langle\psi ({\bf r})\rangle , \\
\nt ({\bf r}, {\bf r}^{'}) & =\nt^{\ast} ({\bf r^{'}}, {\bf r})=\langle\psi^{+} ({\bf r})\psi ({\bf r}^{'})\rangle -\Phi^{*} ({\bf r})\Phi ({\bf r}^{'}), \\
\mt ({\bf r},{\bf r}^{'}) & = \mt ({\bf r^{'}},{\bf r})=\langle\psi ({\bf r})\psi ({\bf r}^{'})\rangle -\Phi ({\bf r})\Phi ({\bf r}^{'}),
\ea
\label{eq1}
\end{equation}
where $\psi ({\bf r})$ and $\psi^{+} ({\bf r})$ are the boson destruction and creation field operators satisfying the usual commutation rules
$[\psi ({\bf r}),\psi^{+} ({\bf r^{'}})]=\delta({\bf r}-{\bf r^{'}})$, $[\psi ({\bf r}),\psi ({\bf r^{'}})]=[\psi^{+}({\bf r}),\psi^{+} ({\bf r^{'}})]=0$. 
These quantities completely characterize the gaussian density operator and, in our langage, are akin to an order parameter and to one-body correlation functions. The local limit of the latter are just the noncondensate density $\lim_{{\bf r}^{'}\to {\bf r}}\nt ({\bf r}, {\bf r}^{'})=\nt ({\bf r})$ and the anomalous density $\lim_{{\bf r}^{'}\to {\bf r}}\mt ({\bf r}, {\bf r}^{'})=\mt ({\bf r})$. Let us consider a second quantized hamiltonian $H$ of the form: 
\be
H=\int_{\bf r} \psi^{+} ({\bf r})H^{\rm {sp}}({\bf r})\psi ({\bf r})+
\demi \int_{{\bf r},{\bf r^{'}}}\,\psi^{+} ({\bf r})\psi^{+} ({\bf r^{'}})V({\bf r},{\bf r^{'}})\psi ({\bf r^{'}})\psi ({\bf r}) ,
\label{eq11}
\end{equation}
where $V({\bf r},{\bf r^{'}})$ is the interaction potential and $H^{\rm {sp}}({\bf r})$ the single particle hamiltonian
\be
H^{\rm {sp}}({\bf r})=-{\ds\hbar ^{2}\over\ds 2m}\, \Delta_{\bf r} +V_{\rm {ext}}({\bf r})-\mu .
\label{eq111}
\ee
$V_{\rm {ext}}({\bf r})$ and $\mu$ are respectively the external trapping field and the chemical potential. The stationarity conditions of the Balian-V\'en\'eroni action-like functional lead to a set of dynamical equations which couple the various densities: 
\be
\ba{rl}
i\hbar \dot{\Phi} ({\bf r})& = H^{\rm {sp}}({\bf r})\Phi ({\bf r})\\
&+\int_{{\bf r^{'}}} \, V({\bf r},{\bf r^{'}})\left[|\Phi ({\bf r^{'}})|^2 \Phi ({\bf r})+\Phi^{\ast}({\bf r^{'}})\mt ({\bf r},{\bf r^{'}})+\Phi ({\bf r^{'}})\nt ({\bf r},{\bf r^{'}})+\Phi ({\bf r})\nt ({\bf r}^{'},{\bf r^{'}})\right], \\ 

i\hbar \dot{\nt} ({\bf r},{\bf r^{'}})  & =\left[H^{\rm {sp}}({\bf r})-H^{\rm {sp}}({\bf r^{'}})\right] \nt({\bf r},{\bf r^{'}})\\
&+\int_{{\bf r^{''}}} \, V({\bf r^{'}},{\bf r^{''}})\left[a({\bf r^{''}},{\bf r^{'}})\nt ({\bf r},{\bf r^{''}})+a({\bf r^{''}},{\bf r^{''}})\nt ({\bf r},{\bf r^{'}})+b({\bf r^{'}},{\bf r^{''}})
\mt^{\ast} ({\bf r^{''}},{\bf r})\right] \\
&-\int_{{\bf r^{''}}} \, V({\bf r},{\bf r^{''}})\left[a({\bf r},{\bf r^{''}})\nt ({\bf r^{''}},{\bf r^{'}})+a({\bf r^{''}},{\bf r^{''}})\nt ({\bf r},{\bf r^{'}})+b^{\ast}({\bf r},{\bf r^{''}})
\mt ({\bf r^{''}},{\bf r^{'}})\right], \\

i\hbar \dot{\mt}({\bf r},{\bf r^{'}})  & =\left[H^{\rm {sp}}({\bf r})+H^{\rm {sp}}({\bf r^{'}})\right] \mt({\bf r},{\bf r^{'}}) \\
&+\int_{{\bf r^{''}}} \, V({\bf r^{'}},{\bf r^{''}})\left[a({\bf r^{''}},{\bf r^{'}})\mt ({\bf r},{\bf r^{''}})+a({\bf r^{''}},{\bf r^{''}})\mt ({\bf r},{\bf r^{'}})+b({\bf r^{'}},{\bf r^{''}})
(\nt^{\ast} ({\bf r},{\bf r^{''}})+\delta ({\bf r}-{\bf r^{''}}))\right] \\
&+\int_{{\bf r^{''}}} \, V({\bf r},{\bf r^{''}})\left[a({\bf r^{''}},{\bf r})\mt ({\bf r^{'}},{\bf r^{''}})+a({\bf r^{''}},{\bf r^{''}})\mt ({\bf r},{\bf r^{'}})+b({\bf r},{\bf r^{''}})
\nt ({\bf r^{''}},{\bf r^{'}})\right], 
\ea
\label{eq3}
\ee
where 
$$
a({\bf r},{\bf r^{'}})=a^{\ast}({\bf r^{'}},{\bf r})=\langle\psi^{+} ({\bf r})\psi ({\bf r}^{'})\rangle = \nt ({\bf r},{\bf r^{'}})+\Phi^{\ast} ({\bf r}){\Phi} ({\bf r^{'}})
$$ 
and 
$$
b({\bf r},{\bf r^{'}})=b({\bf r^{'}},{\bf r})=\langle\psi ({\bf r})\psi ({\bf r}^{'})\rangle = \mt ({\bf r},{\bf r^{'}})+\Phi ({\bf r}){\Phi} ({\bf r^{'}}) .
$$ 
In the case of a contact potential $V({\bf r},{\bf r^{'}})=g\delta({\bf r}-{\bf r^{'}})$, the equations (\ref{eq3}) read 
\be
i\hbar \dot{\Phi} ({\bf r}) = \left[H^{\rm {sp}}({\bf r})+g(\nc ({\bf r})+2\nt({\bf r}))\right]\Phi ({\bf r})+g\mt ({\bf r})\Phi^{\ast} ({\bf r}) ,
\label{eq4}
\ee
\be
\ba{rl}
i\hbar \dot{\nt} ({\bf r},{\bf r^{'}})   =&\left[\left(H^{\rm {sp}}({\bf r})+2gn({\bf r})\right)-\left(H^{\rm {sp}}({\bf r^{'}})+2gn({\bf r^{'}})\right)\right] \nt({\bf r},{\bf r^{'}})\\
& +g\left[b({\bf r^{'}},{\bf r^{'}})\mt^{\ast} ({\bf r},{\bf r^{'}})-b^{\ast}({\bf r},{\bf r})\mt ({\bf r},{\bf r^{'}})\right] ,
\ea
\label{eq44}
\ee
\be
\ba{rl}
i\hbar \dot{\mt}({\bf r},{\bf r^{'}})  =&\left [\left(H^{\rm {sp}}({\bf r})+2gn({\bf r})\right)+\left(H^{\rm {sp}}({\bf r^{'}})+2gn({\bf r^{'}})\right)\right] \mt({\bf r},{\bf r^{'}}) \\
& +g\left[b({\bf r^{'}},{\bf r^{'}})(\delta ({\bf r}-{\bf r^{'}}) +\nt^{\ast}({\bf r},{\bf r^{'}}))+b({\bf r},{\bf r})\nt({\bf r},{\bf r^{'}})\right] ,
\ea
\label{eq444}
\ee
where $\nc ({\bf r})=|\Phi ({\bf r})|^{2}$ is the condensate density and $n({\bf r})=\nc ({\bf r})+\nt ({\bf r})$ the total density. For convenience, we have omitted the time dependence. We notice that even for a contact potential, the equations (\ref{eq44}) and (\ref{eq444}) are still complicated. Indeed, the presence of second order derivatives (in $H^{\rm {sp}}$) and the $\delta$ term in (\ref{eq444}) do not allow for simple local limits. This is precisely the term which misteriously disappear in the derivation of Chernyak {\it et al.}\cite{CHER03} using the coherent state representation. We will see in section 3 that this term is responsible for the divergences while its structure allows for a correct renormalization of the coupling constant.

On the other hand, this set of equations was also derived by \cite{GRIF96, PROUK96, KOH02} in a local form by a quite different approach. In our case however, these equations can be related to the well-known HFB-BdG equations only in the quasi-homogeneous case. Indeed,  upon setting
\be
\ba{rl}
\mt ({\bf r},{\bf r^{'}})&=-\demi\sum_k (1+2n_k )\left(U_k({\bf r})V_k({\bf r^{'}})+U_k({\bf r^{'}})V_k({\bf r})\right), \\
\nt  ({\bf r},{\bf r^{'}})&=\sum_k \left(n_k U_k^{\ast}({\bf r})U_k({\bf r^{'}})+(1+n_k)V_k^{\ast}({\bf r})V_k({\bf r}^{'})\right), 
\ea
\label{eq45}
\ee
where $U_k$ and $V_k$ are linearly independent space functions and $n_k$ is the occupation probability of the mode $k$ which is given at equilibrium by the Bose-Einstein distribution function, we readily get from (\ref{eq44}) and (\ref{eq444})
\be
\ba{rl}
i\hbar \dot{U_k} ({\bf r})   =& \left [H^{\rm {sp}}({\bf r})+2gn({\bf r})\right ]U_k ({\bf r})-g b({\bf r},{\bf r}) V_k^{\ast} ({\bf r}),  \\
i\hbar \dot{V_k} ({\bf r})   =& \left [H^{\rm {sp}}({\bf r})+2gn({\bf r})\right ]V_k ({\bf r})-g b({\bf r},{\bf r}) U_k^{\ast} ({\bf r}) ,
\ea
\label{eq46}
\ee
which, together with (\ref{eq4}) form the so-called (generalized) time dependent HFB-BdG equations \cite{MOR, TT97, GRIF96}. Hence, our equations can be considered as a variational extension of the time dependent HFB-BdG equations for non-homogeneous situations.

Among the interesting properties of the TDHFB equations, we can quote the conservation of the energy and of the total particle number. 
Moreover, the ''Heisenberg'' parameter, which is defined as
\be
I ({\bf r},{\bf r^{'}})=\int_{{\bf r^{''}}}\, \left(\delta ({\bf r}-{\bf r^{''}})+2\nt({\bf r},{\bf r^{''}})\right)
\left(\delta ({\bf r^{''}}-{\bf r^{'}})+2\nt({\bf r^{''}},{\bf r^{'}})\right)-4\int_{{\bf r^{''}}}\, \mt^{*}({\bf r},{\bf r^{''}})\mt({\bf r^{''}},{\bf r^{'}}) ,
\label{eq5}
\ee
and which can be written in the more readable compact form
\be
I ({\bf r},{\bf r^{'}})=\left(\delta ({\bf r}-{\bf r^{'}})+2\nt({\bf r},{\bf r^{'}})\right)^2-4\left |\mt ({\bf r},{\bf r^{'}})\right |^2 ,
\label{eq55}
\ee
is also a constant of motion at zero temperature. Indeed, for $T=0$, $I({\bf r},{\bf r^{'}})=\delta ({\bf r}-{\bf r^{'}})$ and 
$\int_{{\bf r^{''}}}\, \nt^{*}({\bf r},{\bf r^{''}})\mt({\bf r^{''}},{\bf r^{'}})=\int_{{\bf r^{''}}}\, \mt({\bf r},{\bf r^{''}})\nt({\bf r^{''}},{\bf r^{'}})$. 
One can easily check that these properties remain true during the evolution if they are initially fulfilled. This conservation 
law is not however new since it is intimately related to the unitary evolution of the single particle density matrix 
{\cite{BM05, BV, BF99}.

The equation (\ref{eq444}) involves an explicitly diverging term (for ${\bf r}={\bf r^{'}}$) which should be regularized. To this end, one may absorb this diverging expression in a redefinition of the thermal average\cite{MOR, HUTCH, GI04} or, as performed by \cite{KB11, BCB99}, one may use the pseudo-potential method\cite{HU87} to renormalize the coupling constant. A more rigorous approach is the $\Lambda$-potential method first discussed in\cite{Pr11, Pr04}. Neverttheless, these approaches suffer from conceptual difficulties in that they are based on substraction schemes, which are not compatible with the variational technique. Another way of proceeding, which seems to us much more interesting, since it allows to remain in the variational framework, is to resort to a better representation which would ''wash out'' these singular terms. One of the best candidates is the Wigner representation\cite{TT97, RS80, SV00}. To this end, let us define the Wigner transform for any function $A({\bf r},{\bf r^{'}})$ as 
\be
A({\bf r},{\bf r^{'}})=\int_{\bf k} A_W ({\ds {\bf r}+{\bf r^{'}}\over\ds 2},{\bf k})\exp{[-i{\bf k.(r-r^{'})]}},
\label{eq6}
\ee
and its inverse
\be
A_W({\bf R},{\bf k})=\int_{\bf r} A ({\bf R}+{\bf r}/2,{\bf R}-{\bf r}/2)\exp{(i{\bf k.r})},
\label{eq61}
\ee
where $\int_{\bf k}={\ds1\over\ds (2\pi)^3}\int d^3 {\bf k}$. In the limit ${\bf r}={\bf r^{'}}$, the equations (\ref{eq44}) and  (\ref{eq444}) take the form
\be
\ba{rl}
i\hbar \dot{\nt}_{W}({\bf r},{\bf k})  = & -{i\ds\hbar ^{2}\over\ds m}\,{\bf k}.\nabla_{{\bf r}}\nt_{W}({\bf r},{\bf k}) \\
& +g\left[(\Phi^2 ({\bf r})+\mt ({\bf r}))\mt_{W}^{\ast} ({\bf r},{\bf k})-({\Phi^{\ast}}^2 ({\bf r})+\mt^{\ast} ({\bf r}))\mt_{W} ({\bf r},{\bf k})\right], 
\ea
\label{eq62}
\ee
\be
i\hbar \dot{\mt}_{W}({\bf r},{\bf k})  = 2\left[H^{\rm {sp}} +2gn({\bf r})+{\ds\hbar^{2}k^2\over\ds 2m}\right]\mt_{W} ({\bf r},{\bf k})
+g (\Phi^2 ({\bf r})+\mt ({\bf r}))(1+2\nt_{W}({\bf r},{\bf k})).
\label{eq63}
\ee

In the following sections, we will provide the static solutions to these equations both at zero and finite temperature. This will not only yield quantitative estimates of the anomalous effects, but also lead to some important remarks concerning the behaviour of the anomalous density with respect to the dimensionality of the problem. 

\disregard{

Before proceeding further, let us briefly return to the conservation of the total particle number. It is an easy task to show now that the total 
density obeys the continuity equation
\be
{\ds\partial n({\bf r})\over\ds \partial t}+ \nabla .{\bf J}({\bf r}) =0,
\label{eq64}
\ee
where the total current density is given by
\be
{\bf J}={\ds\hbar\over\ds 2mi}\left[{\Phi^{\ast}\nabla\Phi - \Phi\nabla\Phi^{\ast}}\right]
+\int\,{\ds d^3{\bf k}\over\ds (2\pi)^3} {\ds\hbar {\bf k}\over\ds m}\,\nt_W ({\bf r},{\bf k}), 
\label{eq65}
\ee
and as expected has two contributions coming from the condensate ${\bf J_c}$ and the non condensate ${\bf J_{nc}}$ respectively. 
From a hydrodynamical point of view, upon introducing the condensate velocity ${\bf v_c}$ (also called superfluid) and the non condensate 
velocity ${\bf v_{nc}}$, one comes up with a quite natural expression for the current density at finite temperature:
\be
{\bf J}=\nc {\bf v_c} + \nt {\bf v_{nc}} .
\label{eq655}
\ee
The condensate velocity, defined as usual by ${\bf v_c}=\hbar/m\, {\nabla }\theta$ where $\theta$ is the phase of the condensate field $\Phi$, 
is irrotational. Although we can proceed in the same way for the non condensate velocity, which will lead to an irrotational flow, we see that 
according to (\ref{eq65}), ${\bf v_{nc}}$ is a superposition of free particle velocities ($\hbar {\bf k}/m$) which are locally irrotational and conservatives. 
This witnesses a profound difference between the condensate and the non condensate components of the gas.

First of all, one may notice that the conservation of the 
Heisenberg parameter (\ref{eq55}) can be better visualized here. Indeed, the Wigner transform
\be
I_W ({\bf r},{\bf k})=\left(1+2\nt_W ({\bf r},{\bf k})\right )^2 -4 \left |\mt_W ({\bf r},{\bf k})\right |^2 ,
\label{eq66}
\ee
satisfies the continuity equation
\be
{\ds\partial I_W\over\ds \partial t}+ \nabla .{\bf J}_I =0,
\label{eq67}
\ee
where the current density
\be
{\bf J}_I={\ds\hbar {\bf k}\over\ds m} I_W
+{\ds\hbar \over\ds m}\left [4{\bf k}|\mt_W|^2 + {\ds \mt_W^{\ast}\nabla\mt_W - \mt_W\nabla\mt_W^{\ast}\over\ds i}\right ], 
\label{eq68}
\ee
is also made of two pieces which represent respectively the normal current ${\bf J_n}=(1+2\nt_W )^2 \,\hbar {\bf k}/m$
and the anomalous current ${\bf J_{an}}=|\mt_W |^2 {\bf v_{an}}$, where the ''anomalous'' velocity is also proportional to 
the gradient of the phase of the anomalous density and is therefore irrotational.
In this form, this conservation equation reveals the importance of the anomalous density which, if neglected, would inevitably 
lead to inconsistencies.

}

\setcounter{equation}{0}
\section{Anomalous Effects at $T=0$}

In the static case, and for a neutral gas, one may consider without loss of generality that $\Phi$, $\mt$ and $\nt$ are real space functions. The equation (\ref{eq62}) translates into 
\be
{\bf k}.\nabla_{{\bf r}}\nt_{W}({\bf r},{\bf k})=0, 
\label{eq70}
\ee
which means in particular that there is no current associated with the non condensate and that the condensate current is therefore conservative. On the other hand, the equations (\ref{eq4}) and (\ref{eq63}) yield
\be
0  = \left[H^{\rm {sp}}+g(\nc +2\nt +\mt )\right]\Phi({\bf r}) ,
\label{eq71}
\ee
\be
0  = 2\left (H^{\rm {sp}} +2gn({\bf r})+{\ds\hbar^{2}k^2\over\ds 2m}\right)\mt_W ({\bf r},{\bf k})+ g(\nc +\mt )(1+2\nt_W ({\bf r},{\bf k})).
\label{eq711}
\ee
Next, one may use the gradient expansion method\cite{TT97, RS80, SV00} to show that $\nt_W$ and $\mt_W$ write simply
\be
\ba{rl}
2\mt_W ({\bf r},{\bf k})&=-(1+2n_k)\sinh{\sigma ({\bf r},{\bf k})}, \\
1+2\nt_W  ({\bf r},{\bf k})&=\phantom{-}(1+2n_k)\cosh{\sigma ({\bf r},{\bf k})},  
\ea
\label{eq72}
\ee
where we recall that $n_k$ is the Bose-Einstein distribution function at equilibrium. The Wigner transform of the Heisenberg parameter $I_W$ 
(\ref{eq55}) takes the familiar form
\be
\sqrt{I_W}=1+2n_k =\coth{\left(\beta\epsilon_k /2\right )},
\label{eq73}
\ee
$\beta$ being the inverse temperatures. The quantities $\epsilon_k$ are the s.p. energies obtained by solving the stationary Gross-Pitaevskii equation (\ref{eq71}):
\be
\epsilon_k={\ds\hbar ^{2} k^2\over\ds 2m}+V_{\rm {ext}}-\mu +g(n +\nt +\mt).
\label{eq733}
\ee
Corrections to the expressions (\ref{eq72}-\ref{eq73}) will measure how far the gas is from homogeneity\cite{TT97}.

The spectrum (\ref{eq733}) is manifestly gapless. Indeed, in the homogeneous long wavelength limit, the Hugenholtz-Pines theorem\cite{HP} is automatically satisfied (see the end of this section). In contrast, in the standard HFB framework, one has to slightly modify the equations in order to obtain a gapless spectrum (see e.g.\cite{MOR, HUTCH, GRIF96, GR}). Similarly, in the $\Lambda$-potential approach\cite{Pr11, Pr04}, a zero gap condition is imposed in order to render the theory consistent and this leads to the correct equation of state.

In order to analyze the anomalous effects, let us derive workable expressions for $\nt$ and $\mt$. To this end, we follow the method depicted in \cite{SV00}. The key point shown there is that, up to order $\hbar^2$, the Wigner transform of the product of two s.p. operators is the product of their Wigner transforms\cite{TT97, RS80}. The equation (\ref{eq71}) may thus be written formally in a classical form
\be
\left (\mu -H_{cl}\right)\Phi=0,
\label{eq74}
\ee
with the self-consistent hamiltonian $H_{cl}=p^2/2m+V_{\rm {ext}}+g(n +\nt +\mt)$. The local momentum $p$ must satisfy the on-shell condition 
$p\equiv p_0 ({\bf r})=\sqrt{2m (\mu -V_{\rm {ext}}-g(n +\nt +\mt))}$. Using (\ref{eq72}), we get the intermediate result
\be
\tanh{\sigma({\bf r},{\bf k})}={\ds g (n_c+\mt ) \over\ds
{\ds p_0^2\over\ds 2m}+V_{\rm {ext}}-\mu +2gn+{\ds \hbar^2 k^2\over\ds 2m} } ,
\label{eq744}
\ee
which yields the desired expressions for $\nt$ and $\mt$:
\be
\mt =-\demi \int\,{\ds d^3{\bf k}\over\ds (2\pi)^3} 
{\ds g (n_c+\mt ) 
\over\ds
\sqrt{ \left({\ds \hbar^2 k^2\over\ds 2m}+ 2gn_c\right)\left({\ds \hbar^2 k^2\over\ds 2m}- 2g\mt \right)}
}
\coth{(\beta\epsilon_k /2)},
\label{111}
\ee
\be
\nt=\demi \int\,{\ds d^3{\bf k}\over\ds (2\pi)^3} 
\left [-1+
{\ds 
{\ds \hbar^2 k^2\over\ds 2m}+  g (n_c -\mt ) 
\over\ds
\sqrt{ \left({\ds \hbar^2 k^2\over\ds 2m}+ 2gn_c\right)\left({\ds \hbar^2 k^2\over\ds 2m}- 2g\mt \right)}
}
\coth{(\beta\epsilon_k /2)}
\right].
\label{112}
\ee
What is remarkable on these expressions is that the denominator is clearly not of the Bogoliubov type. This is the price to be paid in order to maintain not only the consistency of our variational approach but also the gapless nature of the spectrum. To recover a Bogoliubov-type spectrum, one may set 
$\mt =0$ in the integrand, which is reminiscent to the so-called Popov approximation. But this will be shown to be highly inconsistent in a variational framework.

It is also important at this level to notice that the on-shell condition is somewhat equivalent to the local density approximation in the sense that the effect of the confining potential is only present in the temperature dependent term.

Furthermore, since the previous expressions can be generalized to any dimension $D$ by simply making the replacement 
$\int{\ds d^3{\bf k}\over\ds (2\pi)^3}\to \int{\ds d^D{\bf k}\over\ds (2\pi)^D}$, one can see that they are both free from IR divergences. On the other hand, while $\nt$ is finite at high momenta, we can clearly see that $\mt$ is UV divergent for $D=2$ and $D=3$. It is however finite in the one-dimensional case. 

In the latter case and at zero temperature, a typical length scale (the interaction length) is $a_1=\hbar^2/mg$. A straightforward calculation then yields:
\be
\nt/n_c=\frac{\sqrt{\gamma}}{2\pi}
\sqrt{1-|\mt|/n_c}\left(\hbox{arctanh}\,\frac{1-\sqrt{|\mt |/n_c}}{\sqrt{1-|\mt |/n_c}}-\frac{1-\sqrt{|\mt |/n_c}}{\sqrt{1-|\mt |/n_c}}\right)
,
\label{113}
\ee
\be
|\mt|/n_c=\frac{\sqrt{\gamma}}{4\pi}
\sqrt{1-|\mt|/n_c}\left(\log{(1+\sqrt{1-|\mt /n_c|})}-\log{\sqrt{|\mt /n_c|}}\right)
,
\label{114}
\ee
where we have introduced the dimensionless gas parameter $\gamma =1/\sqrt{a_1n_c}$ which measures the ratio between the interparticle length $1/n_c$ and the interaction length $a_1$. These simplified expressions are obtained under the assumption that $|\mt|/n_c <<1$. In the general case, they will yield the elliptic functions $\bf K$ and $\bf F$. Using this last assumption, we may further simplify the equations (\ref{113}) and (\ref{114}) to get:
\be
\nt/n_c\simeq\frac{\sqrt{\gamma}}{2\pi}\log{(1/\sqrt{|\mt| /n_c})}
,
\label{115}
\ee

\be
|\mt|/n_c\simeq\frac{\sqrt{\gamma}}{4\pi}
\log{(1/\sqrt{|\mt| /n_c})}
.
\label{116}
\ee
We note on these expressions that the anomalous fraction is half the quantum depletion. Although they are both small, it would be hazardous even at zero temperature to neglect the anomalous average while maintaining the non condensate density. Moreover, the equation (\ref{116}) does not admit solutions unless $\gamma <<1$ which is the the weak coupling condition $a_1n_c>>1$ \cite{GS03}. In this case, the explicit solutions may be written as:
\be
{\ds |\mt|\over\ds n_c}\simeq\demi {\ds\nt\over\ds n_c}\simeq\sqrt{\gamma}(1-\sqrt{\gamma}).
\label{117}
\ee
However, the next to leading correction of order $\gamma$ is not relevant since it seems to be related to three-body losses which are not considered here\cite{SHL}. Hence, up to order $\sqrt{\gamma}$, $\nt$ and $\mt$ are of the same order and scale as $\sqrt{\gamma}$.

In order to extend the calculation to more than 1D, one must handle carefully the UV logarithmic (2D) and linear (3D) divergences that appear in the integral (\ref{111}).

Let us begin by the 2D case. First of all, the expression (\ref{112}) can easily be evaluated to yield 
\be
\nt = \frac{mg}{4\pi\hbar^2}n_c \left(1+\frac{\mt}{n_c}\right){\ds q^2\over\ds 1+\sqrt{1-q^4}},
\label{118}
\ee
where $q={\ds n_c+\mt \over\ds n_c-\mt}$ is very close to 1 at low temperatures. On the other hand, the integral (\ref{111}) must be regularized. Upon introducing a cutoff $K$, we obtain for high momenta 
\be
\mt = - {\ds \left(\frac{mg}{2\pi\hbar^2}\right)\log{(2K/K_0)}\over\ds 1+\left(\frac{mg}{2\pi\hbar^2}\right)\log{(2K/K_0)}}\, n_c ,
\label{119}
\ee
where $K_0^2={\ds 4mg\over\ds\hbar^2}n_c$. We may now renormalize the coupling constant by setting $g_R = g \log{(2K/K_0)}$ and introduce the dimensionless gas parameter $\gamma =mg_R/2\pi\hbar^2$ to obtain a finite expression for $\mt$:
\be 
\mt = -{\ds \gamma\over\ds 1+\gamma} \, n_c .
\label{120}
\ee
It should be emphasized that this renomalization of the coupling constant is analogous to what is obtained in \cite{Pr11} (see equation (50) of this reference). Furthermore, our cutoff $K$ is just the parameter $\Lambda$ introduced there. The final expressions for $\mt$ and $\nt$ (replacing in the latter $g$ by $g_R$ and setting $q\simeq 1$) read
\be 
\frac {|\mt|}{n_c}\simeq \frac {\nt}{n_c} = {\ds \gamma\over\ds 1+\gamma},
\label{121}
\ee
which show that $\nt$ and $\mt$ are also of the same order but now scale as $\gamma$.

In 3D, although the calculations are mainly the same, their outcomes are quite different. Indeed, we obtain for $\nt$ and $\mt$ the expressions
\be
\nt = \frac{8}{3\sqrt{\pi}}\left[a(n_c+\mt )\right]^{3/2},
\label{122a}
\ee
\be
\mt = -aK (n_c+\mt ),
\label{122b}
\ee
where $K$ is a high momentum cutoff and $a$ is the ''bare'' s-wave scattering length, which satisfies $aK=-\mt /(n_c+\mt)$. This is the right renormalization condition since upon redefining the coupling constant by
\be
g_R={\ds g\over\ds 1+aK},
\label{123}
\ee
we get $g_R=g(1+\frac{\mt}{n_c})$ which redefines both the scattering length $a\to a_R=mg_R/4\pi\hbar^2$ and the noncondensate density
\be
\nt = \frac{8}{3\sqrt{\pi}}\left[a_R n_c\right]^{3/2}.
\label{124}
\ee
The previous renormalization condition was also derived in \cite{Pr11, Pr04} based on the $\Lambda$-potential and by \cite{MOR, HUTCH} using a somewhat 
ad-hoc prescription. In this sense, the approaches developed in the previous references are compatible although they seem quite different. It is however remarkable that (\ref{124}) can be directly deduced from (\ref{122a})by setting $\mt =0$, which is also compatible with (\ref{122b}) if we set $a=\sum_{s=1}\gamma_s/K^s$. In this case, we get the finite part of $\mt$ 
\be 
\mt =- \frac {\gamma_1}{1+\gamma_1}n_c.
\label{125}
\ee
Hence, unlike the 1D and 2D cases, $\mt$ and $\nt$ are not of the same order of magnitude. The former takes values compatible with 0, as in the HFB-Popov formalism. 

Finally, one may derive the equation of state in each case using the Hugenholtz-Pines theorem which relates the chemical potential to the diagonal and off-diagonal self-energies $\hbar\Sigma_{11}$ and $\hbar\Sigma_{12}$ by $\mu =\hbar\Sigma_{11}-\hbar\Sigma_{12}$. The latter are given by
\be
\left\{
\ba{rl}
\hbar\Sigma_{11} &=2T_D n ,\\
\hbar\Sigma_{12} &=T_D (n_c-\mt) ,
\ea
\right.
\label{126}
\ee
where $T_D$ is the D-dimensional $T$-matrix. In the case of a contact potential, $T_D$ is just the bare coupling constant $g$.

\setcounter{equation}{0}
\section{Anomalous Effects at Finite Temperature}
In order to compute the finite temperature corrections, one may rewrite (\ref{112}) in the form:
\be
\nt=\nt_0+\demi \int\,{\ds d^D{\bf k}\over\ds (2\pi)^D} 
{\ds 
{\ds \hbar^2 k^2\over\ds 2m}+  g (n_c -\mt ) 
\over\ds
\sqrt{ \left({\ds \hbar^2 k^2\over\ds 2m}+ 2gn_c\right)\left({\ds \hbar^2 k^2\over\ds 2m}- 2g\mt \right)}
}
\big (\coth{(\beta\epsilon_k /2)}-1\big).
\label{127}
\ee
where, for low $T$, the first term may be approximated by its value at $T=0$, hence the subscript 0. The second term may be computed as follows. At very low temperatures, the integral is dominated by the behavior of its integrand for $k=k_0=p_0/\hbar$, $p_0$ being the local momentum given by (\ref{eq74}). Upon setting $\delta\nt =\nt -\nt_0$, we obtain:
\be
\delta\nt\simeq\frac{S_D\Gamma(D/2)}{4\pi^{D/2}}  
{\ds 
{\ds \hbar^2 k_0^2\over\ds 2m}+  g (n_{c0} -\mt_0 ) 
\over\ds
\sqrt{\big({\ds \hbar^2 k_0^2\over\ds 2m}+ 2gn_{c0}\big)\big({\ds \hbar^2 k_0^2\over\ds 2m}- 2g\mt_0}\big)
}
\lambda^{-D}\,e^{-\beta\left(V_{\rm {ext}}-\mu+g(n_0 +\nt_0 +\mt_0)\right)},
\label{128}
\ee
where $S_1=1$, $S_2=2\pi$, $S_3=4\pi$, $\Gamma$ is the Euler function and $\lambda$ is the thermal De Broglie wavelength. The subscript 0 in the densities means that they are evaluated at zero temperature. Proceeding in the same way for $\mt$, we get
\be
\delta\mt\simeq\frac{S_D\Gamma(D/2)}{4\pi^{D/2}}  
{\ds 
g (n_{c0} +\mt_0 ) 
\over\ds
\sqrt{\big({\ds \hbar^2 k_0^2\over\ds 2m}+ 2gn_{c0}\big)\big({\ds \hbar^2 k_0^2\over\ds 2m}- 2g\mt_0}\big)
}
\lambda^{-D}\,e^{-\beta\left(V_{\rm {ext}}-\mu+g(n_0 +\nt_0 +\mt_0)\right)}.
\label{129}
\ee
The previous expressions show that the finite temperature corrections although being exponentially small are of the same order for both $\mt$ and $\nt$. Our conclusions concerning the order of magnitude of these two quantities remain therefore valid at finite temperature.

One may further simplify these expressions owing to the equation of state and to the results obtained in section 3. The main point here is that the temperature dependence of the corrections, beside the exponential factor, is $T^{D/2}$.

\setcounter{equation}{0}
\section{Conclusions}

In this paper, we have been mainly interested in obtaining quantitative estimates of the so-called anomalous average for bose condensed gases. 
Indeed, owing to its importance to account for many-body effects, it was not clear for us what is the status of this quantity especially in the HFB formalism.

To this end, we preferred to use from the beginning a consistent formalism. The time-dependent variational principle of Balian and V\'en\'eroni appeared as the best candidate. Indeed, not only does it lead to approximate dynamical equations by a judicious choice of the trial spaces, but it maintains also the consistency of the approach along the whole calculational process. In this context, we obtained first a set of time-dependent equations for the order parameter and the one-body correlation functions, which were shown to generalize the HFB equations in the non-local, non-homogeneous case. Moreover, using the Wigner representation, we were able to take the correct local limit, showing that the anomalous density and the non condensate density are the local limits of the previous correlation functions. This demonstrates in particular that the HFB-BdG equations are only valid in the quasi-homogenous case. Hence, our formalism allows to analyze the anomalous effects in more general situations.   

It is to be recalled that we have previously performed a similar analysis in the low momentum regime and obtained quite interesting results about the anomalous density. The question that naturally arises was whether these results remain true when we take into account the high lying modes. This question becomes crucial at finite temperature.

Our analysis shows that the anomalous density and the non condensate density are of the same order in one and two space dimensions at zero and finite temperature. Therefore, neglecting only the former is a quite hazardous approximation. In 3D however, the situation is much more subtle and requires a careful analysis. A preliminary result is that the anomalous density is effectively small compared to the non condensate density.

Moreover, using a renormalization scheme compatible with our variational approach, we obtained in a more reliable and rigorous way, finite expressions for the anomalous density and the non condensate density similar to what is obtained in the literature. 

Many unanswered questions remain to be examined. One of the most important is whether the previous results depend on the nature of the interactions. Indeed, the contact interaction that we adopted in this paper is the simplest and care should be taken especially for long range interactions. We plan to publish in the near future a similar analysis for dipolar gases.

Moreover, it has been shown elsewhere that the HFB approximation consist in fact of a resummation of ladder diagrams\cite{Pr11} which opens the door to the well known speculations about the relationship between the variational and the perturbative approaches\cite{NE}. 

The time dependent renormalization is also of great importance. In our variational formalism, we feel that this can be done more easily since we have shown elsewhere\cite{BM98} that the time dependent mean field equations are renormalizable once they have been renormalized initially.

We would like to thank G. Shlyapnikov, L. Pricoupenko and M. V\'en\'eroni for many enlightening discussions and kind hospitality.

\end{document}